\documentclass[3p,times,procedia]{elsarticle}
\usepackage{nupha_ecrc}

\volume{00}

\firstpage{1}

\journalname{Nuclear Physics A}

\runauth{}

\jid{nupha}

\jnltitlelogo{Nuclear Physics A}

\usepackage{amssymb}

\usepackage[figuresright]{rotating}

\begin{document}

\begin{frontmatter}



\dochead{XXVIIIth International Conference on Ultrarelativistic Nucleus-Nucleus Collisions\\ (Quark Matter 2019)}

\title{Feasibility studies of conserved charge fluctuations in
Au-Au collisions with CBM}


\author{Subhasis Samanta (for the CBM Collaboration)}
\address{School of Physical Sciences, National Institute of Science Education and Research,
HBNI, Jatni-752050, India}

\begin{abstract}
We present the CBM physics performance study for measurements of the
higher order cumulants of the net-proton multiplicity distributions.
These observables are proxy for net-baryon fluctuations and are
commonly used to study the phase structure of QCD phase diagram.
The simulation is done for $Au+Au$ collision at 
beam kinetic energy $E_{lab} = 10$ AGeV.
The cumulants of net-proton distributions have been calculated at 
midrapidity ($|\Delta y| =1$)
for the transverse momentum range $0.2 < p_T < 2.0$ GeV/c.
The centrality dependence of cumulants of net-proton upto order four is presented. 
The efficiency and detector effects are corrected using unfolding techniques.
This work shows that the higher order cumulants of net-proton can be measured using 
the CBM detector.
\end{abstract}




\end{frontmatter}


\section{Introduction}
Substantial theoretical
as well as experimental efforts world-wide have been devoted
to investigate properties of matter under extreme conditions.
According to the Lattice Quantum Chromodynamics (LQCD) calculation there should be 
a smooth crossover transition between hadronic phase and quark-gluon plasma 
(QGP)
at high temperature $T$ and zero baryon chemical potential $\mu_B$~\cite{Aoki:2006we}. On the other hand, 
various QCD based models ~\cite{Asakawa:1989bq} predict
a first-order phase transition at low $T$ and high $\mu_B$.
Hence, there must be a critical point (CP) at high $T$ and non-zero 
$\mu_B$ where the first-order phase transition line ends.
Several heavy-ion collision experimental program 
world-wide have been devoted to the investigation of QCD
matter over a wide range of $T$ and $\mu_B$. 
Heavy ion collision experiments at the Large
Hadron Collider (LHC), CERN and Relativistic Heavy Ion Collider (RHIC) at BNL
are presently investigating QCD matter at high $T$
and small $\mu_B$ region of the phase diagram.
To search the QCD CP a Beam Energy Scan (BES) program \cite{Adam:2020unf}
is ongoing at RHIC. 
A similar energy and system size scan by the NA61/SHINE~\cite{Davis:2019mlt} collaboration at CERN SPS is also
in progress.
The HADES experiment \cite{Agakishiev:2015bwu}
at GSI, Darmstadt is investigating a medium at the very large $\mu_B$.
In future, the Nuclotron-based Ion Collider Facility
(NICA) at JINR, Dubna and
the Compressed Baryonic Matter (CBM) experiment \cite{Ablyazimov:2017guv} at the Facility for 
Antiproton and Ion Research (FAIR) at GSI will also study nuclear matter at large 
$\mu_B$ region of the phase diagram. 
The CBM experiment at SIS100 synchrotron will offer the
opportunity to study the nuclear matter with very high precision data
at the center of mass energy range $\sqrt{s_{NN}} = 2.7 - 4.9$ GeV
for $Au+Au$ collisions~\cite{Ablyazimov:2017guv}.

Fluctuations and correlations of conserved charges are believed to be 
important observables to 
search of the phase transition and the CP of the QCD phase diagram.
Here conserved charges are those which are conserved in strong
interaction like baryon number, strangeness number or electric charge.
Experimentally, net-charge $N_q = N_q^+-N_q^-$ multiplicity distribution
is measured in a finite acceptance on an event-by-event basis. 
The $n^{th}$ order central moment of the distribution is defined as
$\langle (\delta N_q)^n\rangle=\langle (N_q-\langle N_q\rangle)^n \rangle,$
where $\langle N_q\rangle$ is the mean value of the distribution.
The higher order moments $\langle (\delta N_q)^3\rangle$, $\langle (\delta N_q)^4\rangle$
are expected to be sensitive to the matter properties in the vicinity of the CP \cite{Stephanov:2008qz}.
To characterize a distribution, cumulants are used. Cumulants are related
to the central moments by the following relations:
\begin{equation}
 C_1 = \langle N_q \rangle, ~~C_2 = \langle (\delta N_q)^2 \rangle,~~ C_3 = \langle (\delta N_q)^3 \rangle,~ \mathrm{and}~  C_4 = \langle (\delta N_q)^4 \rangle -3 \langle (\delta N_q)^2 \rangle^2.
\end{equation}
The cumulants upto order four are studied in the present work.
Cumulants are connected to the theoretically calculable
susceptibilities of the conserved charges by the 
relation $C_n = VT^3\chi_q^n$ \cite{Karsch:2010ck, Gupta:2011wh},
where $\chi_q^n$, $V$ and $T$ are the $n^{th}$ order susceptibility,
volume and temperature of the system, respectively.

\section{Analysis detail}
In this performance study, 
net-proton (which are used as a proxy of net-baryon)
fluctuation for $Au + Au$ collision at the beam kinetic energy $E_{lab} = 10$ AGeV
have been calculated.
Simulations have been performed within the CbmRoot
framework. Five million minimum bias events generated by UrQMD model~\cite{Bleicher:1999xi}
are transported through the CBM detector 
setup using GEANT3 Monte-Carlo.
The list of CBM detector subsystems simulated for this study includes:
Micro Vertex Detector (MVD),
Silicon Tracking System (STS), Ring Imaging Cherenkov Detector (RICH), 
Transition Radiation Detector (TRD) and Time-of-Flight Detector (TOF).
The lab pseudorapidity range covered by CBM detector system is
$1.5 < \eta < 3.8$.
The MVD gives the collision vertex of the event and the STS provides the momentum
information of the charged particle tracks.
\begin{figure}[htb]
\centering
\includegraphics[width=0.4\textwidth]{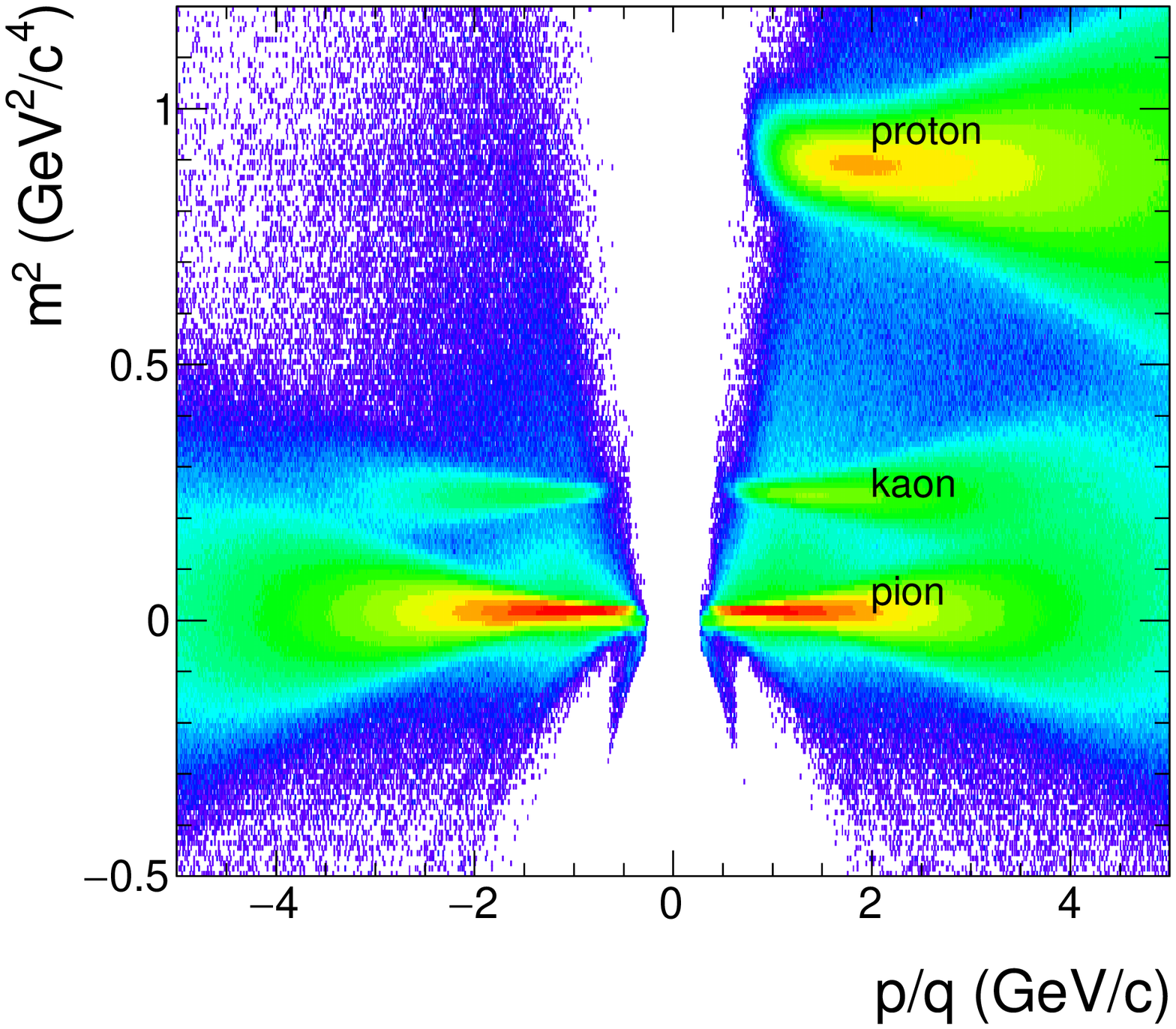}
\includegraphics[width=0.4\textwidth]{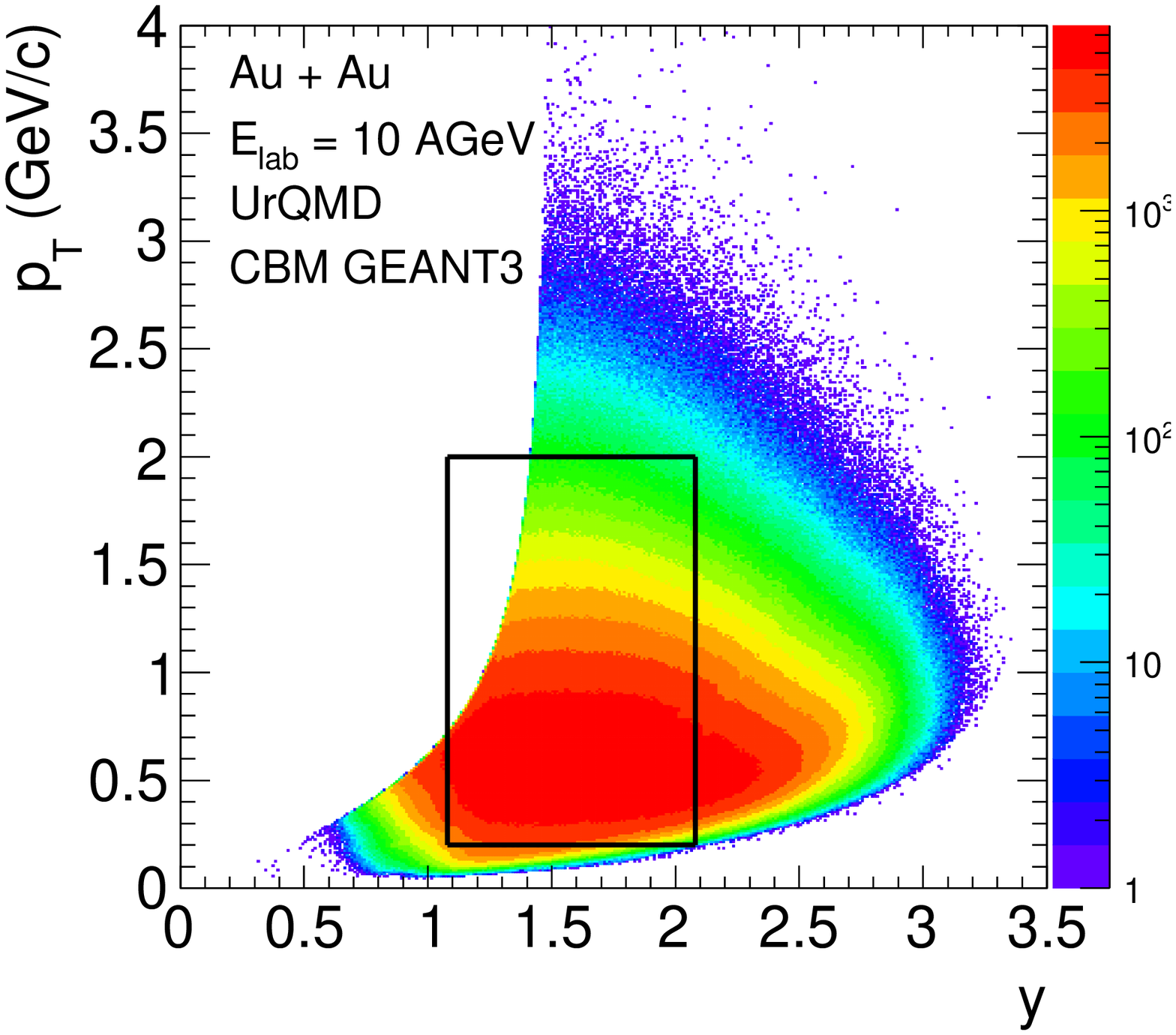}
 \caption{(Left) Distribution of $m^2$ as measured by the TOF detector as a function of $p/q$.
(Right) $p_T$ vs $y$ distribution of protons and anti-protons in $Au+Au$ collisions at $E_{lab} = 10$
AGeV. Box indicates the region of analysis of net-proton cumulants. 
 }
 \label{fig:m2}
 \end{figure}

\subsection{Particle identification}
For the purpose of particle identification, TOF detector has been used.
The TOF detector provides flight time ($t$) of a particle
inside the detector. By knowing the length of the track ($L$), we
can calculate velocity ($\beta = v/c = L/tc$, $c$ is the velocity of light in vacuum),
which depends on mass ($m$) and the momentum ($p$) of the particle. Hence 
one can calculate the relation of $m^2$ with $p$ using $m^2 = p^2(1/\beta^2-1)$.
Left panel of Fig. \ref{fig:m2} shows distribution of $m^2$ and $p/q$, where $q$ is the charge of the particle.
For positive values of $p/q$ one can see three well separated bands which 
corresponds to
$\pi^+$, $K^+$ and proton, respectively. 
Anti-particles have a negative value of $p/q$. At the CBM energies significantly 
smaller number of 
anti-protons compared to that of protons is expected.
Tracks with $m^2$
between 0.6 and 1.2 GeV$^2$/c$^4$ are identified as protons or anti-protons
with purity more than 96\%.
This figure shows
that clean proton identification is possible and hence one can
study the net-proton (proxy for net-baryon) higher order moments using
CBM detector.

%
\subsection{Centrality estimation}
The collision centrality is determined using the
reconstructed charged particle multiplicity 
measured with STS. The charged particles are selected using
$m^2$ less than 0.4 GeV$^2$/c$^4$ which exclude the
protons and anti-protons for multiplicity measurements and avoids self correlation 
effects. The analysis is done
in nine centrality bins 0-5\%, 5-10\%, 10-20\%, 20-30\%, 30-40\%, 40-50\%, 50-60\%, 60-70\% and 70-80\%.
Centrality bin 0-5\% corresponds to most central
whereas 70-80\% corresponds to the most peripheral collision events.
\section{Results}
\begin{figure}[htb]
\centering
\includegraphics[width=0.4\textwidth]{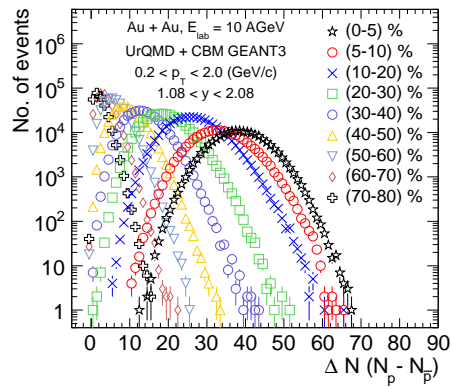}
\caption{Distributions of net-proton multiplicities in different centrality bins.}
\label{fig:np_dis}
\end{figure}

In this work we calculate cumulants up to the order of four net-proton ($\Delta N = N_p - N_{\bar{p}}$)
multiplicity distribution. The protons and anti-protons are counted within the transverse 
momentum range $0.2 < p_T < 2.0$ GeV/c and unit rapidity ($|\Delta y =1|$) 
window at mid-rapidity relative to the beam rapidity $y_{beam} = 1.58$.
The $y-p_T$ acceptance for protons and anti-protons selection in $Au+Au$
collision at $E_{lab} = 10$ AGeV is shown in 
right panel of Fig. \ref{fig:m2}.
Efficiency of protons and anti-protons detection in different collision centrality varies between 62 \% to 46 \%.
Figure \ref{fig:np_dis} shows the reconstructed event-by-event
(without efficiency correction)
net-proton multiplicity
distributions for different centrality classes.
The finite width of a centrality class may cause volume fluctuations within
this class.
The cumulants need to be corrected for such effects~\cite{Luo:2013bmi}. The centrality bin
width corrected cumulants are calculated as $ C_n = w_r C_{n,r}$ where
$C_{n,r}$ and
 $w_r$ are respectively cumulant and the fraction of event in $r^{th}$ multiplicity bin i.e.
$w_r = n_r\sum_r n_r$. The $n_r$ is the number of events 
and cumulant in $r^{th}$ multiplicity bin.

The measured cumulants are also affected by the reconstruction efficiencies of the 
proton and anti-protons. We have
corrected for this and other detector effects using the method of
unfolding~\cite{DAgostini:1994fjx}. For
this purpose we use the RooUnfoldBayes algorithm which
is based on the Bayes theorem. The
measured and true number of particles are related by the
relation $ y = R \cdot x$,
where $y$ and $x$ are the measured and true distribution and $R$ is a response
matrix.
We have divided the simulated data
set in two halves where first half is used to reconstruct the
response matrix and the second half of the data set is used for
the analysis. The response matrix is used to get the
true number of protons and anti-protons from the information of 
measured (reconstructed) protons and anti-protons.
\begin{figure*}[htb]
\centering
\includegraphics[width=0.28\textwidth]{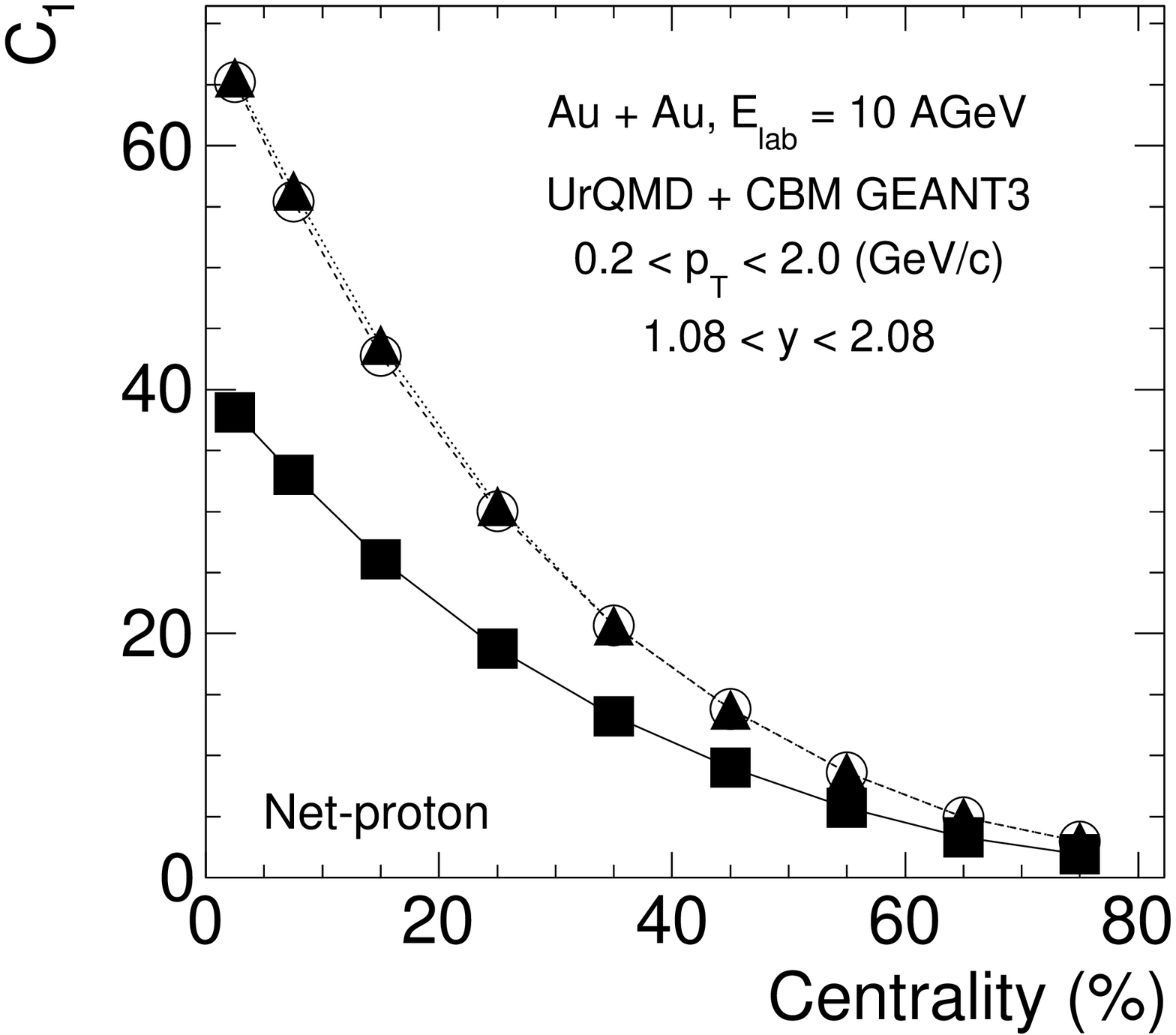}
\includegraphics[width=0.28\textwidth]{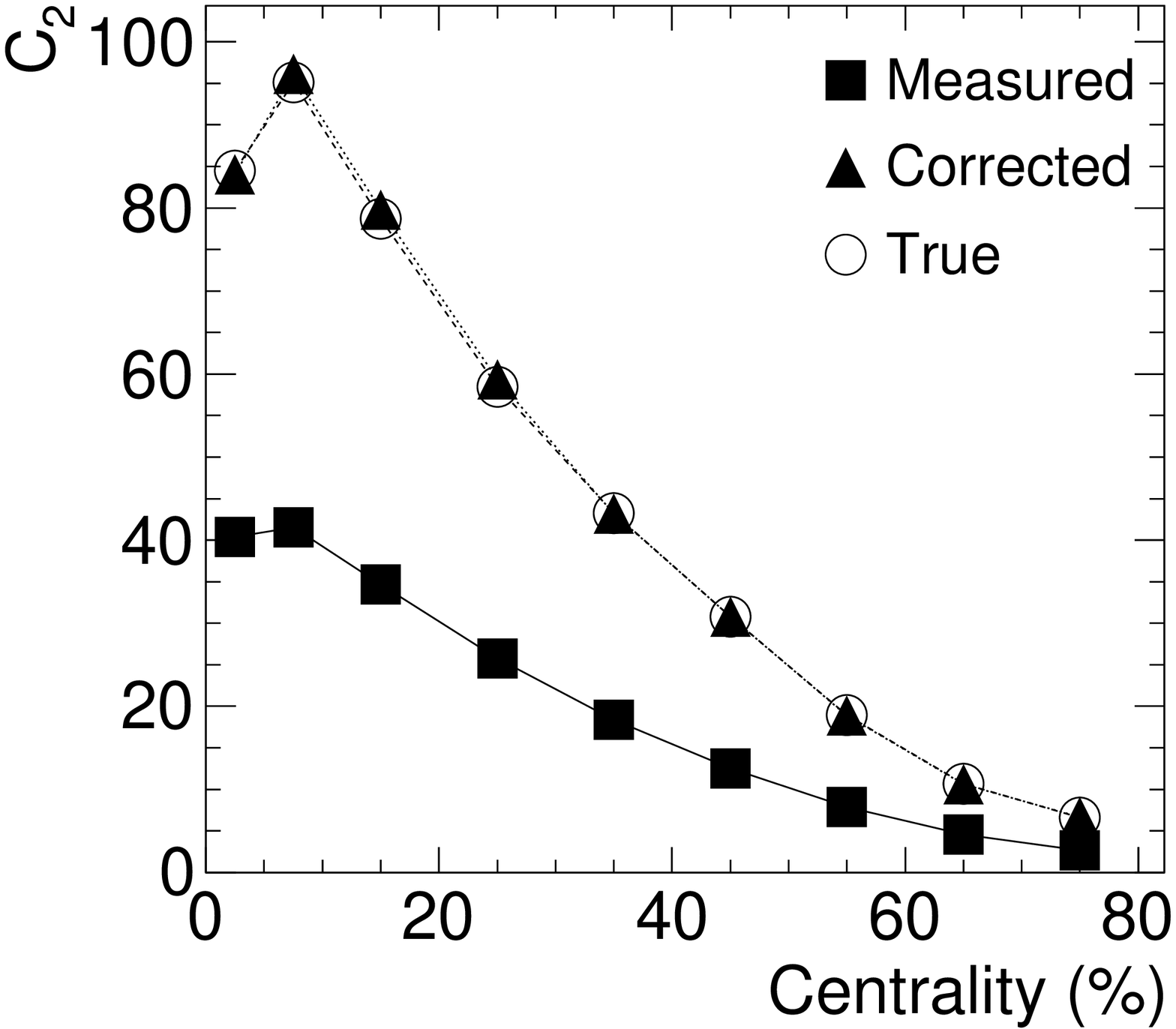}\\
\includegraphics[width=0.28\textwidth]{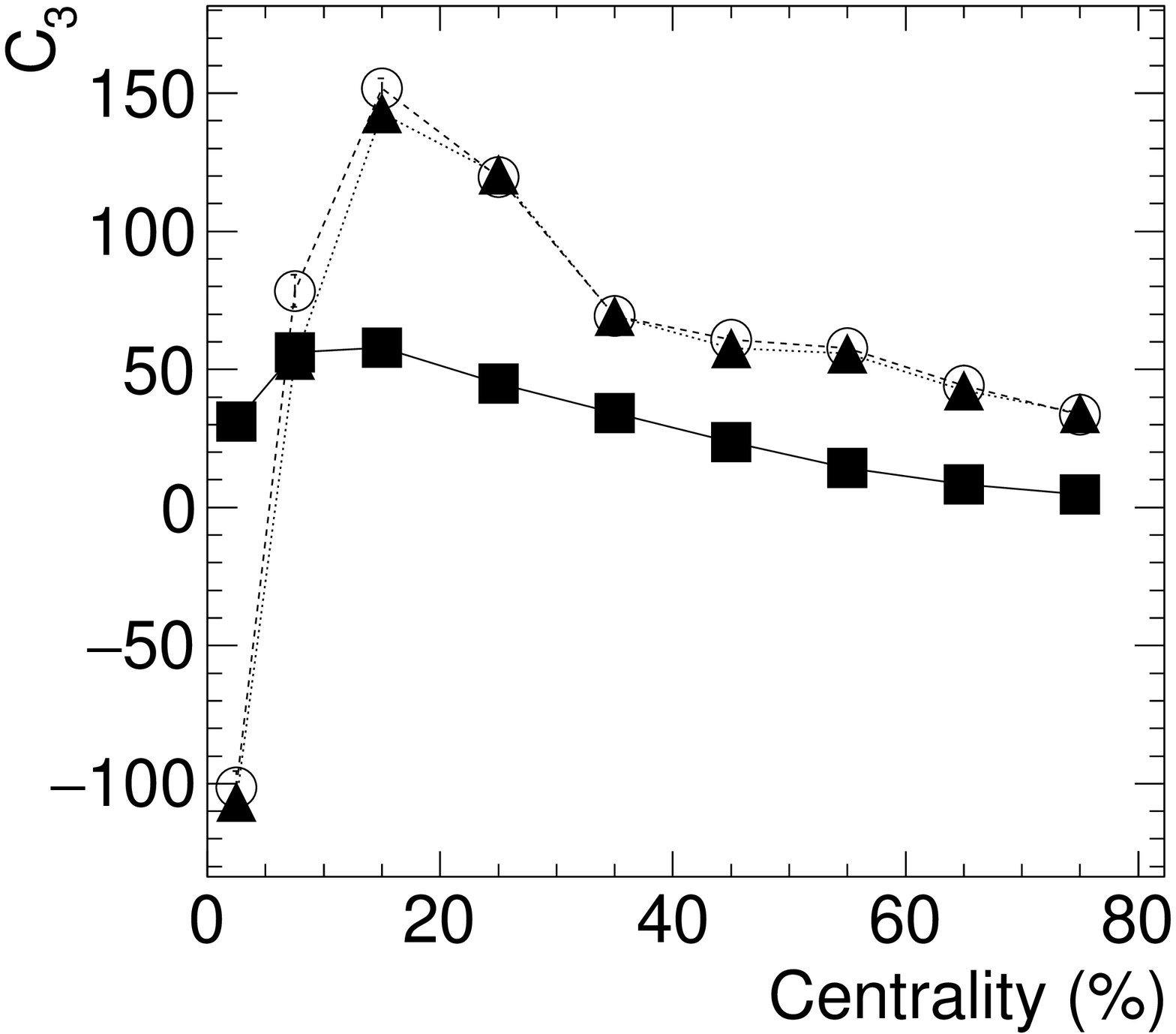}
\includegraphics[width=0.28\textwidth]{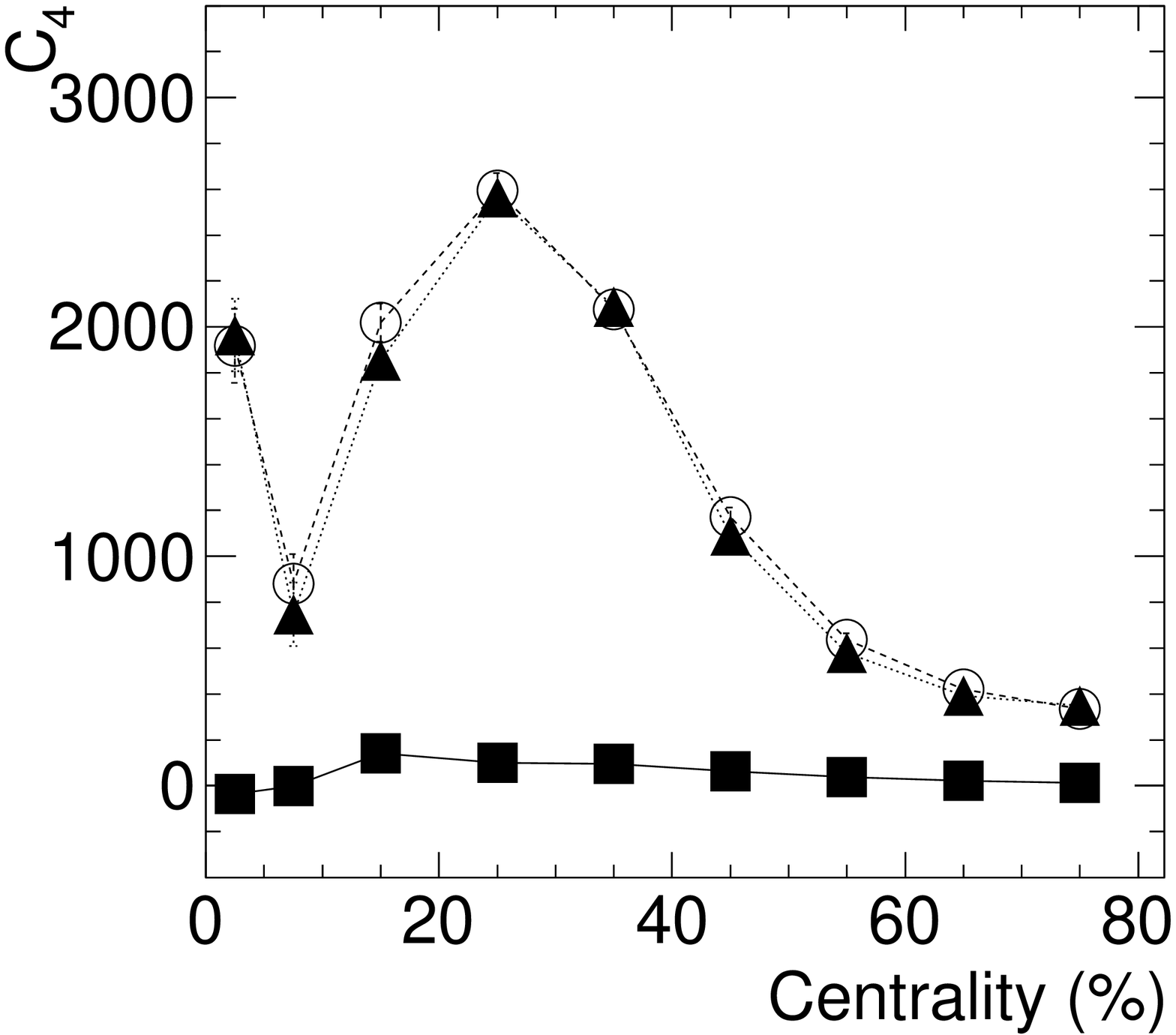}
\caption{The centrality dependence of cumulants of net-proton distributions up to order four.}
\label{fig:cn}
\end{figure*}

Centrality dependence of cumulants of net-proton multiplicity distribution 
up to the order of four
have been shown in Fig. \ref{fig:cn}.
The solid squares correspond
to the centrality bin width corrected cumulants calculated from reconstructed 
net-proton distributions whereas open circles show the
generated net-proton cumulants.
The
solid triangles show the cumulants obtained after
the application of unfolding procedure. Within statistical
uncertainties unfolding method is able to reproduce 
the true value of the cumulants.
The volume of the produced QCD matter (or the number of 
participants) decrease from central to peripheral collisions.
Hence, decrease of cumulants from central to peripheral collisions are expected,
which is indeed observed except for few centrality bins 
($C_2$ in 0-5 \%, $C_3$ in 0-5 \% and 5-10 \%, $C_4$ in 0-5 \%, 5-10 \% and 10-20 \%).
The error bars shown on the data point are statistical only
which are estimated using delta theorem approach~\cite{Kendall}.

In summary, we have performed the physics performance study of net-proton fluctuations 
for CBM detector setup. 
It demonstrates the feasibility of doing higher moments measurements of 
net-proton distributions using CBM detector.
In future we plan to analyse cumulants
up to order six with large statistics and other energies of
SIS100. 
Similar analysis
is also ongoing for the net-charge
and will be done for net-kaon and mixed cumulants.

\noindent
\textbf{Acknowledgement}\\
I acknowledge financial support from Department of Atomic Energy (DAE), 
Government of India.
I also gratefully acknowledge the support provided by the Indo-FAIR Co-ordination
Center at Bose Institute (BI-IFCC), Kolkata, India for this work and attending the conference.









\end{document}